\documentstyle[12pt,twoside,fleqn,espcrc1]{article}

\newcommand{\AmS}{{\protect\the\textfont2
  A\kern-.1667em\lower.5ex\hbox{M}\kern-.125emS}}
\newcommand{\be}{\begin{eqnarray}}
\newcommand{\ee}{\end{eqnarray}}

\renewcommand{\vec}[1]{{\bf #1}}

\input epsf
\newcommand{\grpicture}[1]
{
    \begin{center}
        \epsfxsize=200pt
        \epsfysize=0pt
        \vspace{-5mm}
        \parbox{\epsfxsize}{\epsffile{#1.ps}}
        \vspace{5mm}
    \end{center}
}

\title{ PHYSICS OF HOT AND DENSE QCD }

\author{A. Smilga \address{Dept. of Physics, University of Nantes, 2 rue de la
 Houssiniere, BP 92208 F-44322, Nantes CEDEX 3, France } \address{
ITEP, B. Cheremushkinskaya 25, Moscow 117259, Russia}}

\begin{document}
\maketitle

\begin{abstract}
We give a brief review of modern theoretical
understanding of the physics of $QCD$ at finite temperature and density.
We concentrate in particular on the properties of hadron gas at relatively
small temperatures and discuss in details the physics of the phase transition.
We argue that the phase transition in {\it temperature} is probably absent, but
it can appear with a vengeance when finite density effects are taken into 
account.
We notice also that the temperatures one can expect to reach at the heavy
ion collider at RHIC are not yet high enough for the perturbation theory in 
the QCD coupling constant to work well.
\end{abstract}

\section{Introduction.}
The properties of $QCD$ medium at finite temperature have
been
the subject of intense study during the last 15--20 years.
It was
realized that the properties of the medium undergo a drastic
change as the temperature increases. At low temperatures,
the
system presents a gas of colorless hadron states:  the
eigenstates of the $QCD$ hamiltonian at zero temperature.
When
the temperature is small, this gas is composed mainly of
pions
--- other mesons and baryons have higher mass and their
admixture in the medium is exponentially small $\sim
\exp\{-M/T\}$. At small temperature, also the pion density
is
small --- the gas is rarefied and pions practically do not
interact with each other.

 However, when the temperature increases, the pion density
grows,
the interaction becomes strong, and also other strongly
interacting hadrons appear in the medium. For temperatures
of
order $T \sim$ 150 Mev and higher, the interaction is so
strong that the hadron states do not present a convenient
basis
to describes the properties of the medium anymore, and no
analytic calculation is possible.

 On the other hand, when the temperature is very high, much
higher than the characteristic hadron scale $\mu_{hadr}
\sim$
0.5 Gev, theoretical analysis becomes possible again. Only
in
this range, the proper basis are not hadron states but
quarks
and gluons --- the elementary fields entering the $QCD$
lagrangian. One can say that, at high temperatures, hadrons
get
"ionized" to their basic compounds.
In the $0^{\underline{th}}$ approximation, the
system presents the heat bath of freely propagating colored
particles. For sure, quarks and gluons interact with each
other, but at high temperatures the effective coupling
constant is small $\alpha_s(T) \ll 1$ and the effects due to
interaction can be taken into account perturbatively.
This interaction has the long--distance Coulomb nature,
and the properties of the system are
in
many respects very similar to the properties of the usual
non-relativistic plasma involving charged particles with
weak
Coulomb interaction. The only difference is that quarks and
gluons carry not the electric, but color charge. Hence the
name: {\it Quark-Gluon Plasma} ($QGP$).

 Thus the properties of the system at low and at high
temperatures have nothing in common. A natural question
arises:
What is the nature of the transition from low-temperature
hadron
gas to high- temperature quark-gluon plasma ? Is it a {\it
phase} transition ? If yes, what is its order ?  I want to
emphasize that this question is highly non-trivial. A
drastic
change in the properties of the system in a certain
temperature
range does not guarantee the presence of the phase
transition
{\it point} where free energy of the system or its specific
heat
is discontinuous. Recall that there is no phase transition
between the ordinary gas and the ordinary plasma.

 There are at least 4 reasons why this question is
interesting
to study:
\begin{enumerate}
 \item It is just an amusing
theoretical question.
 \item Theoretic conclusions can be
checked in lattice numerical experiments.  Scores of papers
devoted to lattice study of thermal properties of QCD have
been
published.  \item {\it Some} theoretical results can be confronted with the
results of the experiments done in
{\it laboratory} with heavy ion colliders. In particular, one could expect
to obtain interesting results on  the collider
  RHIC which is now under construction.
\item During the first
second of its evolution, our Universe passed through the
stage
of high-$T$ quark-gluon plasma which later cooled down to
hadron
gas (and eventually to dust and stars, of course). It is
essential to understand whether the phase transition did
occur
at that time. A {\it strong} first-order phase transition
would lead
to observable effects.  We know (or almost know --- the
discussion of this question has not yet completely died
away)
that there were no such transition. But it is important to
understand why.
\end{enumerate}

There is also a related
question
--- what are the properties of relatively cold but very
dense
matter and whether there is a phase transition when the
chemical
potential corresponding to the baryon charge rather than the
temperature is increased. From experimental viewpoint this question is no 
less interesting and important: in the realistic heavy ion experiments we have
{\it both} finite temperature and finite baryon number density.
 Though theoretical predictions are less certain here,   we have acquired 
recently some qualitative understanding of high 
density physics and I am going to discuss it later in my talk.
A limited time of the talk and a limited volume of this contribution did not
allow us to dwell upon many interesting aspects of hot QCD physics in details.
Also the list of references is very incomplete. More detailed discussion can be
found in the review \cite{obzor}.   
  
\section{Temperatures are never high.}
\setcounter{equation}0

The temperature one can hope to achieve in heavy ion collision experiment is 
not too high. For example, the upper bound for the expected temperature in 
RHIC
experiment is about $T_{\rm RHIC} \sim 500 $ MeV \cite{RHICT} and may be twice 
as much at LHC.

Let us ask  whether the system can already be described in terms of
quarks and gluons at such temperatures and can be safely called 
``quark-gluon plasma''. In other words: whether the effective
QCD coupling constant $\alpha_s(T)$ is small enough so that the perturbation 
theory works well ? We will find out that the answer to this question is 
unfortunately
{\it negative} (with some reservations).

Let us consider the basic thermodynamic characteristic of  a finite $T$
hadron system --- its free energy.  In the lowest order, we can neglect
interaction and the problem is reduced to finding the free energy of 
noninteracting quark and gluon gas.
For a pure gluon system, the answer can be written immediately using the 
classical Stephan and Boltzmann result:
 \be
\label{Fg0}
 \frac{F^g}{V} =
- \frac{\pi^2 T^4}{45} (N_c^2 -1)
\ee
where $(N_c^2-1)$ is the number of color degrees of freedom.  The quarks
are fermions and integration over momenta with the Fermi rather than Bose 
distribution gives a somewhat different factor. We have 
 \be
\label{Fq0}
 \frac{F^q}{V} \ = \
 - \frac{7\pi^2 T^4}{180} N_c N_f
\ee
where $N_c N_f$ is the number of degrees of freedom. 
Notice that in our World with $N_c = N_f = 3$, the lowest
order
quark contribution  to the free
energy is roughly 2 times larger than the gluon one. In the diagrammatical 
language, these results  correspond to calculation of one loop 
gluon and quark bubbles at finite T.
Perturbative corrections to this quantity were a subject of intense study 
during last 20 years \cite{ShurD}. 
The first correction $\sim \alpha_s T^4$ comes from two--loop diagrams and 
presents no particular complications. It turns out, however, that starting 
from the 3-loop level, we cannot restrict ourselves with a given order of 
perturbation theory but have to sum over a certain class of so called 
``ring'' graphs.
\footnote{ The reason for that is the bad infrared behavior of the thermal 
graphs which I have no time to discuss here.}
 Breaking of naive perturbation series results in the terms in the expansion
which are not analytic in the coupling constant $\alpha_s$ --- after the term
$\sim \alpha_s$ we have the term $\sim \alpha_s^{3/2}$, the next term is 
 $\sim \alpha_s^{2}$ but also  $\sim \alpha_s^{2} \ln(\alpha_s)$, and the next
term is $\sim \alpha_s^{5/2}$. Anyway, the coefficients of all these terms 
can be found exactly and the result is 
  \be
 \label{Ffull}
F = - \frac{8\pi^2T^4}{45}
\left[ F_0 + F_2 \frac {\alpha_s(\mu)}{\pi} + F_3 \left(
\frac {\alpha_s(\mu)}{\pi} \right)^{3/2} +
\right. \nonumber \\
\left.
F_4 \left( \frac {\alpha_s}{\pi} \right)^{2} +
F_5 \left( \frac {\alpha_s}{\pi} \right)^{5/2} +
O(\alpha_s^3 \ln \alpha_s ) \right]
 \ee
where

\newpage

\be
\label{F0c}
F_0 = 1 + \frac {21} {32} N_f, \ \ F_2 = -\frac{15}4 \left(1 + \frac {5} {12}
 N_f \right), \ \ \ F_3 = 30 \left( 1 + \frac {N_f}6 \right)^{3/2} \ ,
\nonumber \\
F_4 = 237.2 + 15.97 N_f - 0.413 N_f^2 + \frac {135}2
\left( 1 + \frac {N_f}6 \right) \ln \left[ \frac
{\alpha_s}\pi \left( 1 + \frac {N_f}6 \right) \right] 
\nonumber \\
- \frac {165}8 \left(1 + \frac {5} {12} N_f \right)
\left(1 -\frac {2} {33} N_f \right) \ln \frac \mu{2\pi T} \ , \nonumber \\
F_5 = \left(1 + \frac { N_f}6 \right)^{1/2} \left[
-799.2 - 21.96 N_f - 1.926 N_f^2 \right. \nonumber \\
\left. + \frac {485}2 \left(1 + \frac { N_f}6 \right)
\left(1 -\frac {2} {33} N_f \right) \ln \frac \mu{2\pi T}
\right]
 \ee
The expressions (\ref{F0c}) are written for
$N_c =3$. The coefficients like 237.2 are not a result of
numerical integration but are expressed via certain special
functions.
A nice feature of the result (\ref{Ffull}) is its renorm-
invariance.
The coefficients $F_4$ and $F_5$ involve a
logarithmic $\mu$--dependence in such a way that the whole
sum does not depend on the renormalization scale $\mu$.
It is important to note that the term $\sim \alpha_s^{5/2}$ in the expansion
(\ref{Ffull}) is the absolute limit beyond 
which no perturbative calculation is possible. On the level $\sim \alpha_s^3$,
we are running into the so called magnetic mass problem, in other words the
infrared divergences in the graphs become so bad that no recipe for their
resummation can be suggested.

Let us choose $\mu = 2\pi T$ (this is a natural choice,
$2\pi T$ being the lowest nonzero gluon Matsubara frequency
) and $N_f = 3$.
In that case, we have
 \be
 \label{series}
F = F_0 \left[1 - 0.9 \alpha_s + 3.3\alpha_s^{3/2} + (7.1 +
3.5
\ln \alpha_s ) \alpha_s^2 - 20.8 \alpha_s^{5/2} \right]
 \ee
Note a large numerical coefficient at $\alpha_s^{5/2}$. It
is rather troublesome because the correction $\sim
\alpha_s^{5/2}$ overshoots all previous terms up to very
high temperatures and, at temperatures which can  be
realistically ever reached at accelerators, makes    the
whole perturbative approach problematic.

Take $T \sim 0.5 \ {\rm GeV}$. Then $2\pi T \sim 3 \ {\rm
GeV}$
and $\alpha_s \sim 0.2$. (We use a conservative estimate for
$\alpha_s$ following from $\Upsilon$ physics.
Recent
measurements at LEP favor even larger values.). The series
(\ref{series}) takes the form
  \be
  \label{sernum}
 F = F_0[1 - .18 + .3 +.06 - .37 + \ldots]
 \ee
which is rather unsatisfactory. For $T \sim 1$ GeV (the upper bound for the
expected temperature at LHC), the situation is just a little bit better.

We conclude thereby that the hadron matter to be produced at RHIC at LHC 
{\it can}not be called quark--gluon plasma in the technical meaning of this 
word
(a system with weak Coulomb--like interaction). It is a kind of ``no man`s 
land'',
a terrain which is difficult from
the theoretical viewpoint: we cannot describe the system 
neither in terms of hadrons nor in terms of quarks and gluons. {\it Some}
results can, however, be obtained. First, there are numerical results coming
from the lattices. Second, some qualitative theoretical observations on  the
nature and characteristics of the phase transition can be made.

\section{Lukewarm pion gas.}
\setcounter{equation}0
But before discussing the real experimental situation, consider the system 
where the temperature is {\it small} (less than $\sim 100$ MeV) and the baryon
density is zero. {\it Such} system is well feasible to the exact theoretical
analysis.

Let us first remind the well known facts on the dynamics of
$QCD$ at zero
temperature. Consider YM theory with $SU(3)$ color group and
involving $N_f$
massless Dirac fermions in the fundamental representation of
the
group. The fermion part of the lagrangian is
\be \label{Lf}
L_f
= i \sum_f \bar q_f \gamma_\mu {\cal D}_\mu q_f
\ee
where ${\cal
D}_\mu = \partial_\mu - igA_\mu^a t^a$ is the covariant
derivative. The lagrangian (\ref{Lf}) is invariant under
chiral
transformations of fermion fields:
\be
\label{chi}
q_{{\small
L,R}} \rightarrow A_{{\small L,R}}\ q_{{\small L,R}}\ ,
\ee 
where
$q_{{\small L,R}} = \frac 12 (1 \pm \gamma^5)q$ are  flavor
vectors with $N_f$ components and $A_{{\small L,R}}$ are two
different $U(N_f)$ matrices. Thus the symmetry of the
classical
lagrangian is $U_L(N_f) \otimes U_R(N_f)$. Not all N\"other
currents corresponding to this symmetry are conserved in the
full quantum theory. It is well known that the divergence of
the
singlet axial current $ j^5_\mu = \sum_f \bar q_f
\gamma_\mu \gamma^5 q_f$ is nonzero due to anomaly:
$\partial_\mu j_\mu^5  \ \sim \ g^2
\epsilon^{\mu\nu\alpha\beta} G^a_{\mu\nu} G^a_{\alpha\beta}$
Thus the symmetry of quantum theory is $SU_L(N_f) \otimes
SU_R(N_f) \otimes U_V(1)$. It is the experimental fact that
(for
$N_f = 2,3$, at least) this symmetry is broken spontaneously
down to $U_V(N_f)$. The order parameter of this breaking is
the
chiral quark condensate $SU(N_f)$ matrix
  \be
\label{Sigff1}
\Sigma_{ff'}\ =\ < \bar q_{Lf} q_{Rf'}>_0
  \ee
By a proper chiral transformation  it can be
brought
into diagonal form
$\Sigma_{ff'} = - ( \Sigma/  2) \delta_{ff'}$
In the following, the term ``quark condensate'' will be
applied to
the scalar positive quantity $\Sigma$.

For $N_f = 3$, the spontaneous
breaking of chiral symmetry leads to appearance of the octet of
pseudoscalar
Goldstone   states in the spectrum. Of course,
the quarks are not exactly massless in  real $QCD$, and the
mass term is not
invariant with respect to the symmetry (\ref{chi}).
 As a result, in real world we have the octet
of light (but not massless) pseudo-goldstone pseudoscalar
states ($\pi, K, \eta$). But the small mass of
pseudogoldstones
and the large splitting between the massive states of
opposite
parity ($\rho/A_1$, etc.)  indicate beyond reasonable doubts
that the exact chiral symmetry  would be broken
spontaneously in the massless case. As the masses of the
strange
and, especially, of $u$- and $d$- quarks are small
\cite{GLmass}
, the mass term in the lagrangian can be treated as
perturbation.
 E.g. the pion mass satisfies the relation
\be
\label{pimass}
F_\pi^2 M_\pi^2 = (m_u + m_d) \Sigma
\ee
($F_\pi = 93$  Mev is the pion decay constant) and turns to
zero
in the chiral limit $m_{u,d} \rightarrow 0$.

The presence of light pseudogoldstones in the spectrum is of
paramount importance for the physics of low temperature
phase. When
we heat the system a little bit, light pseudoscalar states (
pions
in the first
place) are excited, while $\rho$--meson, nucleon and other
massive degrees
of freedom are still frozen.
We can study {\it analytically} the properties of the system
at low temperatures when the medium presents a rarefied
weakly
interacting gas of pions with low energies.  Their
properties
are described by the effective nonlinear chiral lagrangian
\be
\label{Lchi}
{\cal L} = \frac 14 F_\pi^2 {\rm Tr} \{\partial_\mu
U \partial_\mu U^\dagger\}\ + \ \Sigma {\rm Re\ Tr}\{{\cal
M} U^\dagger\}
\ +\ \ldots
\ee
 where $U\ =\ \exp\{2it^a \phi^a/F_\pi\}$ is the
$SU(N_f)$ matrix ($\phi^a$ are the pseudogoldsone fields),
${\cal M}$ is
the quark mass matrix
 and the dots stand for higher derivative terms
and the terms of higher order in quark masses.
 When the characteristic
energy and the quark masses are small, the effects due to
these
terms are suppressed and a perturbation theory (the {\it
chiral perturbation theory } \cite{CPT} ) can be developed.
The particular nonlinear form of the lagrangian (\ref{Lchi})
which involves 4--pion interaction vertices etc. 
is dictated by the symmetry considerations.

When we switch on
the temperature which is small, only
 the lightest
degrees of freedom, the pions, are excited. Their density is
small and their interaction [described by the effective lagrangian 
(\ref{Lchi})] is weak.

At $T = 0$, we habitually describe the system in terms of  particles 
displaying   themselves as asymptotic
states. In the heat bath, the notion of  asymptotic states (and the notion
of S--matrix)
makes no sense:
due to omnipresent medium the would--be asymptotic states scatter on the 
thermal excitations and do not survive until the collision point. Their
role is taken over
by {\it collective excitations}. This notion can be easily
visualized when
recalling the familiar college physics problem of
propagation of electromagnetic
waves in the medium. The frequency and the wave vector of
such a wave do not
satisfy the vacuum relation $\omega = c|\vec{k}|$ anymore. A
refraction index may
appear.
In plasmas (also in QGP), the dispersion law is modified more essentially and a
mass gap develops.
The amplitude of the classical wave decreases with time due
to dissipative
effects. That means that the wave frequency acquires a
negative imaginary
part (a positive imaginary part would correspond to
instability).

Technically, collective excitations display themselves as
poles in the 
thermal Green's functions 
$<\pi(x) \pi(0)>_T, \ <\bar N(x) N(0)>_T, <\rho(x) \rho(0)>_T $, etc.
\footnote{ the Green`s functions should be {\it retarded} ones, but we will 
not discuss these technicalities here.} The position of the pole
$\omega_{\rm pole}(\vec{k})$ of the Fourrier image of the Green`s function 
is called the {\it dispersive law} of the 
collective excitation with corresponding quantum numbers.

A problem to find the dispersive law for pion collective
excitations is more simple theoretically and was solved first
  \cite{GLT}.
As a result, it was
found that the pion mass [the real part of the pole $\omega(k=0)$] acquires a 
temperature  correction
  \be
 \label{MpiT}
M_\pi^2(T) \ =\ M_\pi^2(0) \left[1 + \frac {T^2}{24 F_\pi^2}
\ + \ldots \right]
  \ee
We see that the pion mass 
slightly increases with temperature, but, if the
quark masses and $M_\pi(0)$ would be zero, pion mass would remain
zero also at finite
temperature. This is easy to understand: massless pions are
Goldstone particles
appearing due to spontaneous breaking of chiral symmetry.
But if the temperature
is small, chiral symmetry is still spontaneously broken and
the massless
goldstones should still be there.

The dispersive laws have not only the real, but also the imaginary part which 
is physically 
associated with the {\it damping} $\zeta(k)$ of the excitations  propagating 
in the
medium and can be observed experimentally as the broading of corresponding 
resonance peaks. For the pions, damping rapidly grows with temperature. The 
average value of the damping [viz.  $\zeta(k)$ integrated over momenta with 
Bose distribution] was found to be \cite{Goity}
 \be
\label{pidam}
<\zeta^\pi>_T  \ \approx \ \frac {T^5}{24 F_\pi^4}
 \ee

Also other hadron states change their
properties when
temperature
is switched on. Nucleons, vector mesons etc. are not easily
excited when the
temperature is  small, but   it makes sense still to study
the problem
of propagation of a massive state in pion gas and find out
how the presence
of the thermal heat bath affects its dispersive properties.
Such a study was
first carried out  for nucleons in \cite{nucl}. The method used was the virial
expansion over the pion density. In the lowest order, the shift of the pole
depends on the amplitude of $\pi N$ scattering which is very well known from
experiment. A brief summary of this study is that
the nucleon mass practically does not vary with the temperature while its
damping grows rapidly. 
At $T \sim
150 \ {\rm MeV}$, the damping is of order $\zeta \sim 100$ MeV and we cannot
reliably calculate and even talk about it at still higher 
temperatures: the 
description of the system in terms of individual hadrons and also in terms of
collective excitations with hadron quantum numbers makes no sense in this
case.

\section{Vector mesons. CERES experiment.}

Unfortunately, the theoretical results for pions and for nucleons 
cannot be directly confronted with experimental data
in heavy ion
collisions. 

We hasten to comment that the pole positions of the
collective excitations
with pion and nucleon quantum numbers are quite physical
quantities and can
be measured in a {\it gedanken} experiment.

Suppose we study the spectrum of invariant masses of two
$\gamma$ emitted
from hot hadronic matter. The spectrum has a sharp peak
associated with the
$\pi^0 \to 2\gamma$ decay. For free pions, the width of the
peak is very
small $\Gamma_\pi^0 \approx 8\ eV$. But {\it thermal} pions
have a finite
width $\Gamma_\pi^T = 2 <\zeta^\pi>_T$.
Also
the maximum of the
resonance is shifted towards larger mass values according to
Eq.(\ref{MpiT}).
Likewise, for nucleons. Proton is believed to decay
eventually, and most
of the decay modes (like
$e^+ \pi^0$) involve hadrons which are stuck within the
hadronic medium
and do not go out undisturbed, but there is also a mode $e^+
\gamma$.
For the argument
sake, let us assume that $e^+$ and $\gamma$ do not interact
with the heat
bath, but only with the detector. Measuring the distribution in
invariant masses
of $e^+\gamma$
and focusing on the ``nucleon resonance'', one could, in
principle, extract
information on the position of the nucleon pole.

Of course, these gedanken experiments are quite unrealistic.
Not even
speaking of the ``nucleon resonance experiment'' which is a
pure science
fiction, also the studying of $2\gamma$ spectrum in  heavy
ion collisions
would provide little information about pion properties in
heat bath. The
matter is that $\pi^0$ lifetime is at least six orders of
magnitude larger
than the lifetime of the fireball produced in the collision
of heavy nuclei
. Pions would decay on flight in vacuum and their mass and
width would be
the same as in any other experiment.

As far as experiment is concerned, the situation is much
better for vector
mesons, especially for the $\rho$--meson. Its vacuum lifetime is
pretty small,
 almost all  $\rho$-s produced in the collision  decay {\it
inside} the
fireball, and measuring the spectra of $e^+e^-$ or
$\mu^+\mu^-$ pairs {\it
can} provide an information about the properties of $\rho$
in hot hadron
medium. Things are not so good with $\omega$ and $\phi$
mesons. Lifetime
of hadronic fireball is estimated to be 25 fm/c or less
\cite{Hung}. It is
of the same order as the $\omega$ lifetime and almost 2
times less than $\phi$
lifetime. A considerable fraction of $\omega$-s and most of
$\phi$-s would
decay outside the fireball.

Theoretically, the problem to find the dispersive law of the $\rho$ - meson
collective excitations is somewhat more complicated than for nucleons
(the main difficulty is that the amplitude of the $\pi \rho$ scattering
in contrast to $T_{\pi N}$ amplitude is not known from experiment), but 
something can be said, however. First, there is an exact theorem that, as is
the case for the nucleon mass, the $\rho$ - meson mass is not shifted to the
order $\sim T^2$  \cite{Dey}. The shift to the order $\sim T^4$ can be 
estimated calculating the amplitude $T_{\pi \rho}$ in the model of
$A_1$ exchange. The results are similar for nucleon case: the mass is
practically not shifted with the temperature (at least, while the temperature
is not so high that the very notion of the $\rho$ - meson collective excitation
loses sense) while the damping (the width of the resonance) grows rapidly.
\footnote{The question is still under discussion. In particular,
the analysis of Ref.\cite{GBrown} displays a rapid decrease of the mass of 
$\rho$ with the temperature. However, this calculation was based on certain 
model assumptions which, 
in our opinion,  do not lead to correct results in this case.}
   
\begin{figure}
\begin{minipage}[t]{80mm}
\grpicture{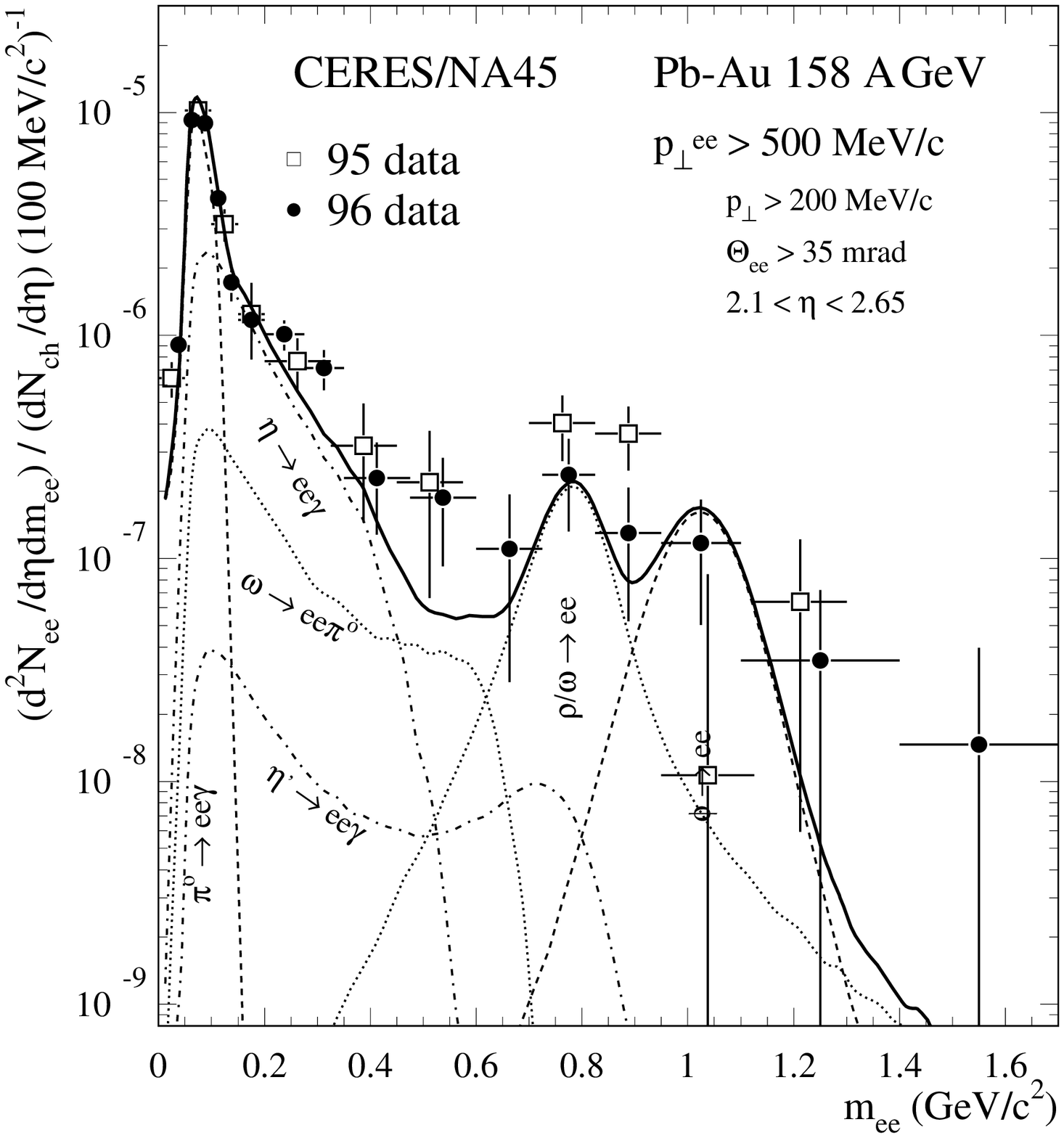}
\end{minipage}
\hspace{\fill}
\begin{minipage}[t]{75mm}
\grpicture{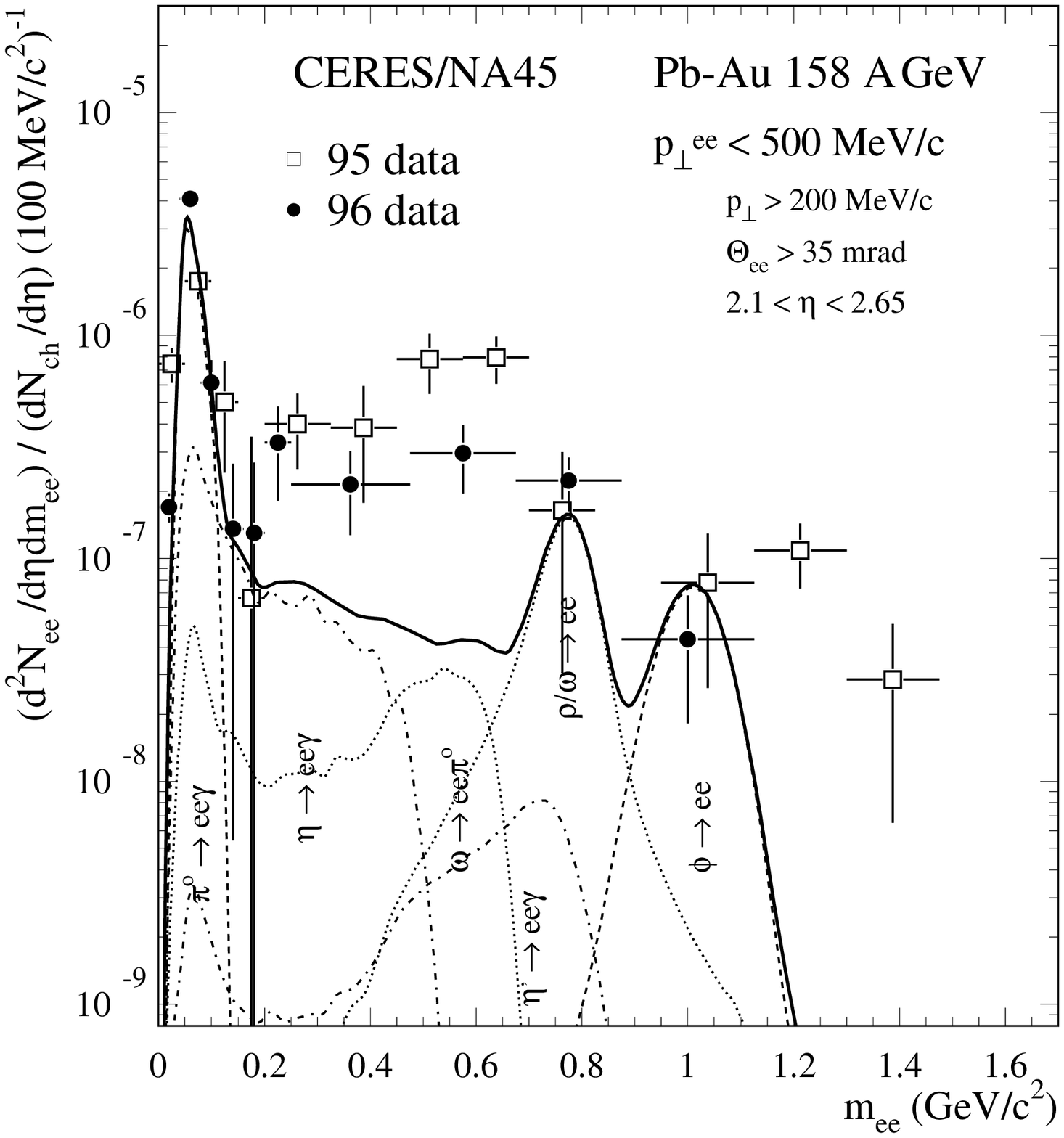}
\end{minipage}

\caption{CERES spectrum of $e^+e^-$ pairs with different transverse momenta.
 A background from individual 
hadron decays is shown. Solid line is the net contribution of all resonances. }
\label{CERES}
\end{figure}

The spectrum of invariant masses of $e^+e^-$ pairs in heavy ion collisions
 has been studied experimentally by CERES collaboration. The {\it excess} of 
events with low invariant masses which cannot be explained if taking into
account only the decay of individual hadron resonances has been found. In Fig.
\ref{CERES} we present the brand new CERES plot \cite{CERES} including the 
separation of data by transverse momenta
of the pair. It was found that the 
whole excess is due to the region of small $p^{e^+e^-}_T $. If imposing the cut
$p^{e^+e^-}_T < 500 \ {\rm MeV/c}$, the signal is strong and 
unequivocal  while, with the cut $p^{e^+e^-}_T >
 500 \ {\rm MeV/c}$, there is no signal at all. It was found also that the 
excess
of low mass $e^+e^-$ pairs correlates with the {\it square} of the hot medium
density.

Obviously, this is a {\it collective} medium effect which  cannot be 
explained in the model with decay of individual resonances. A possible model 
explanation for that effect is the process of $\pi^+\pi^-$ annihilation
  \be
\label{pipian}
\pi^+ \pi^- \ \to \ \rho^*(T, \mu) \ \to \ e^+e^-
 \ee
Just the broadening of the $\rho$ - peak associated with the finite temperature
$T$ and the finite chemical potential $\mu$ can explain qualitatively the 
excess of pairs with low $M_{e^+e^-}$, its quadratic correlation with the 
density and the absence of the effect for large  $p^{e^+e^-}_T $ (pions with
large  $p_T $ are not so abundant in the medium). There is still a problem to
describe the CERES data quantitatively. May be, under the real experimental 
conditions where the temperature about 150 MeV is achieved, one cannot talk
about individual hadrons anymore, and CERES result provides us with the first
glimps of the no man`s land between hadron gas and quark-gluon plasma which
will be studied extensively at RHIC...

\section{Chiral symmetry restoration.}
\setcounter{equation}0

A physical symmetry which is broken spontaneously at zero temperature is 
usually restored when the temperature is high enough (there are some 
rather artificial counterexamples  which we will not discuss here). 
Thereby,
the chiral symmetry should also be restored at high T. At least in {\it 
massless} QCD where the chiral symmetry is the exact symmetry of the lagrangian
this restoration should be associated with the {\it phase transition}. We'll 
come
to this soon, but let ask  first whether something can be said about it by
studying the properties of the system in the low temperature phase where
chiral symmetry is still there and chiral perturbation theory makes sense.

\begin{figure}
 \begin{center}
        \epsfxsize=350pt
        \epsfysize=0pt
        \vspace{-5mm}
       \epsfbox[132 335 566 610]{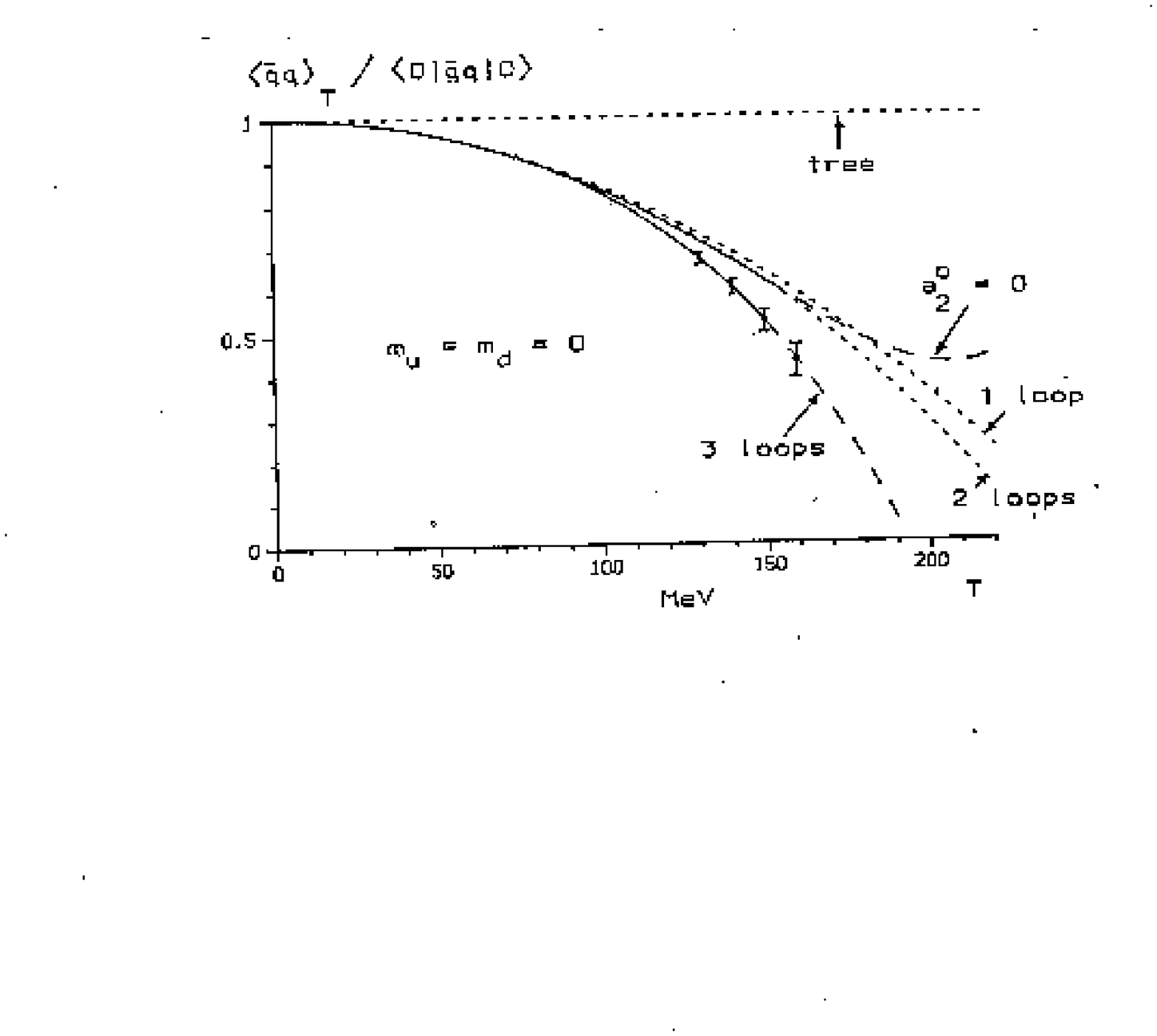}
        \vspace{5mm}
    \end{center}
\caption{Temperature dependence of the condensate in three
loops.}
\label{Gerberl}
\end{figure}

The important result obtained from the chiral analysis is that the chiral
condensate $\Sigma$, the
order parameter associated with chiral symmetry breaking, {\it decreases} 
with temperature. 
The three--loop CPT calculations have been done  in Ref. \cite{Ger}.
The final
 result for the condensate in massless theory in soft pion
approximation
(i.e. when the effects due to $K$ and $\eta$ are disregarded
together with
effects coming from other resonances)
 has a rather simple form

\be \label{qqT}
\Sigma_T = \Sigma_0
\left[ 1 - \frac {T^2}{8F_\pi^2} - \frac {T^4}{384F_\pi^4} -
\frac {T^6}{288F_\pi^6} \ln
\frac \Lambda T + \ldots \right]
  \ee
Here all effects from the higher-derivative term are
described by
the constant $\Lambda$. Experimental data on pseudogoldstone
interactions
give the value
$\Lambda \sim 500 \pm 100$  Mev. The dependence (\ref{qqT})
together with the curves where only the 1-loop correction
$\propto T^2$ or also the 2-loop correction $\propto T^4$
are taken into account
(please, do not put attention to the ``technical'' curve
marked $a^0_2 = 0$)
 are drawn in Fig. \ref{Gerberl}.

The expansion in the parameter $\sim T^2/8F_\pi^2$ makes
sense
when this parameter is small, i.e. when $T \ ^<_\sim\  100-
150$ Mev.
Strictly speaking, one cannot extrapolate the dependence
(\ref{qqT}) for larger temperatures, especially having in
mind
that, at $T > 150$ Mev,  the heat bath includes a
considerable
fraction of other than pion hadron states. But as we know
anyhow
that the phase transition associated  with restoration of chiral
symmetry
should occur, the estimate of the phase transition
temperature
(i.e. the temperature when $\Sigma_T$ hits zero, and, as we'll argue
a bit later,  the transition should be of the second
order at  $N_f = 2$)
based on such an extrapolation is not altogether stupid.
This estimate is
\be \label{est}
T_c \approx 190 \ {\rm Mev}
\ee
or, roughly, one inverse fermi.
A more accurate
treatment which takes into account nonzero $m_{u,d}$ and
also
the presence of other mesons in the heat bath gives
practically
the same estimate for the phase transition temperature
(assuming, of course, that the phase transition is there
which,
 as we will see soon, is probably not the case): nonzero quark masses
smoothen the
temperature dependence of the condensate and push $T_c$ up
while
 excitation of other degrees of freedom sharpen the
temperature dependence
and pulls it down. These
 two effects  practically cancel each other.

However, any calculation based on the soft pion technique fails when the
temperature is of order $T_c$. At such temperatures, we cannot describe the
system in terms of goldstone fields only (and, indeed, an estimate shows that
the contribution of other states in the free energy is not suppressed at 
$T \geq 150 $ GeV).
 Suppose, however, that in the region
$T
\sim T_c$ some other effective lagrangian in Ginzburg-Landau
spirit can be written which depends on the composite
colorless
fields
$\Phi_{ff'} = \bar q_{Rf} q_{Lf'}$. 
A general form of the effective potential which is invariant
under $SU_L(N_f) \otimes SU_R(N_f)$ is
\be
 \label{V3}
V[\Phi] \sim  g_1 {\rm Tr}\{ \Phi \Phi^\dagger\}
+ g_2 ({\rm Tr}\{ \Phi \Phi^\dagger\})^2 
+ g_3{\rm Tr}\{ \Phi \Phi^\dagger \Phi \Phi^\dagger\}
+ g_4 (\det \Phi + \det \Phi^\dagger ) + \ldots
 \ee
(the coefficients may be smooth functions of $T$).  Now look
at
the determinant term. For $N_f = 2$, it is quadratic in
fields
while, for $N_f = 3$, it is cubic in fields and the
effective
potential acquires the structure 
which is characteristic for the systems with first order
phase
transition. A more refined analysis \cite{PW} shows that the
first order
phase transition is allowed also for $N_f \geq 4$, but not
for
$N_f =2$ where the phase transition is of the second order.

Most of the existing lattice data 
indicate that,
at $N_f = 3$, the phase transition is of the first order,
indeed,
while, at $N_f = 2$, it is of the second order.

Up to now, we have concentrated  on studying the  $QCD$ with
massless quarks. But the quarks have nonzero masses: $m_u
\approx$ 4 Mev, $m_d \approx$ 7 Mev, and $m_s \approx$  150
Mev \cite{GLmass}. The
question arises whether the nonzero masses affect the
conclusion on the existence or non-existence and the
properties
of the phase transition.

\begin{figure}
\begin{center}
        \epsfxsize=300pt
        \epsfysize=0pt
       \epsfbox[0 0 320 290]{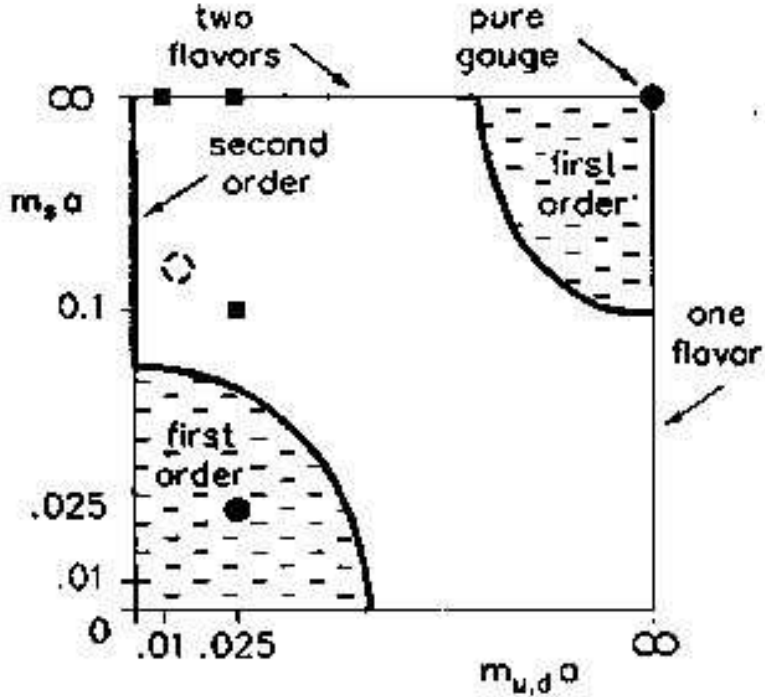}
    \end{center}
\caption{Phase diagram of $QCD$ according to
Ref.\protect\cite{Columb}.
Mass values (measured in the units of the inverse lattice spacing 
$a^{-1}$) for 
which the transition is and is not seen are denoted respectively
by solid circles and squares. Dashed circle corresponds to
the physical
values of masses.}
\label{Columbl}
\end{figure}

Two different
experimental (i.e. lattice) works where
this question was studied are available now, and the results
of these two
studies drastically disagree with each other. The results of
Columbia
collaboration \cite{Columb} display high sensitivity of the
phase
transition dynamics to the values of light quark masses. In
Fig.
\ref{Columbl},
a phase diagram of
$QCD$ with different values of quark masses $m_s$ and $m_u =
m_d$ as drawn in Ref.\cite{Columb} is plotted.

Let us discuss different regions on this plot. When the
quark
masses are large, quarks effectively decouple and we have
pure
YM theory with $SU(3)$ gauge group where the phase
transition appears to be
of the first order (that is also obtained with some independent theoretical
arguments). When all the quark masses are zero, the
phase transition is also of the first order. When masses are
shifted from zero a little bit, we still have a first order
phase transition because a finite discontinuity in energy
and
other thermodynamic quantities cannot disappear at once when
external
parameters (the quark masses) are smoothly changed.

But when all the masses are nonzero and neither are too
small
nor too large, phase transition is absent. Notice the bold
vertical line on the left. When $m_u = m_d = 0$ and $m_s$ is
not
too small, we have effectively the theory with two massless
quarks and the phase transition is of the second order. The
experimental values of quark masses (the dashed circle in
Fig.\ref{Columbl})
 lie close to this line of second order phase transitions
but
in the region where no phase transition occurs. It is the
experimental fact as measured in Ref. \cite{Columb}.

\newpage

\newpage

This statement conforms nicely with a semi-phenomenological
theoretical argument of Ref. \cite{KK} which displays that
even
{\it if} the first order phase transition occurs in QCD, it
is
rather weak. The argument is based on a generalized
Clausius-Clapeyron relation. In college physics, it is the
relation connecting the discontinuity in free energy at the
first-order phase transition point with the sensitivity of
the
critical temperature to pressure. The Clausius-Clapeyron
relation in $QCD$ reads
\be
\label{KK}
  {\rm disc} <\bar q q>_{T_c}\  = \ \frac 1{T_c} \ \frac
{\partial T_c}
{\partial m_q} {\rm disc} \ \epsilon
  \ee
where ${\rm disc}\ \epsilon$ is the latent heat.
The derivative $\frac {\partial T_c}{\partial m_q}$ can the
estimated from theoretical and experimental information of
how
other essential properties of $QCD$ depend on $m_q$ and from
the
calculation of $T$ - dependence of condensate at low
temperature
in the framework of chiral perturbation theory (see
Fig.\ref{Gerberl} and the
discussion thereof).  The
dependence on quark masses is not too weak: $\partial
T_c/\partial m_q \approx
0.9 - 1.0$. From that, assuming
 that the discontinuity in quark
condensate is as large as $<\bar q q>_0$,
we arrive at an estimate
 \be
 \label{Lheat}
{\rm disc}\ \epsilon \ < 0.4 \ {\rm GeV/fm}^3
 \ee
This upper limit for ${\rm disc}\ \epsilon$ should be
compared with the
characteristic energy density of the hadron medium at $T
\sim T_c \sim 190$ MeV.
 The
latter cannot be calculated analytically and we have to rely
on numerical
estimates. The lattice study in \cite{Blum} (for the theory
with 2 massless
flavors) gives a rather large value $\epsilon(T \sim 200\
{\rm MeV}) \
\approx \ 3\ {\rm GeV}/{\rm fm}^3$.
One can safely conclude that the limit (\ref{Lheat})
for the latent heat
 is several times smaller than $\epsilon(T_c)$.
In reality, the discontinuity in $<\bar q q>$ at the phase
transition point
(if it is there)
is, of course, much smaller than the value of the condensate
at zero
temperature. It is seen from the graph in Fig.\ref{Gerberl}.
The condensate
drops significantly still in the region where $K$ and $\eta$
degrees of
freedom are not yet effectively excited. These degrees of
freedom
(and also
$\rho$ and $\omega$ degrees of freedom) become relevant only
at
$T \ ^>_\sim\  150$ MeV when the tendency of the condensate
to drop down is
already well established. Moreover,
as was mentioned earlier in the paragraph
after Eq.(\ref{est}), taking into account these degrees of
freedom
{\it sharpens} the condensate dependence rather than
smoothens it.

Thus latent heat of the first order phase transition in the
theory with 3
massless quarks
must be rather small [significantly smaller than the
estimate (\ref{Lheat})]
 which means that the phase transition is likely to
disappear
under a relatively small perturbation due to nonzero $m_s$.
The numerical
findings of Ref.\cite{Columb} are thereby rather appealing
from theoretical
viewpoint.

\begin{figure}
\grpicture{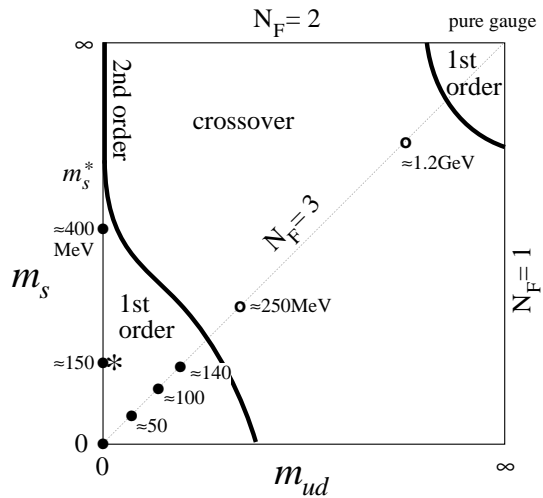}
\caption{Phase diagram of $QCD$ according to
Ref.\protect\cite{Iwas}.
First order phase transition is seen at solid circles and is
not seen at
blank circles.
The star marks out the physical values of quark masses.
The bend of the second order phase transition line
separating the region
with first order phase transition and the region with no
phase transition
reflects some recent theoretical findings \cite{Rajag1}.}
\label{Iwasl}
\end{figure}

However, as far as numerical lattice experiment is concerned, the question is
far from
being  completely resolved by now. A recent lattice study
\cite{Iwas} done
with Wilson rather than Kogut--Susskind fermions displayed
quite a
different picture shown in Fig.\ref{Iwasl}.

What is very surprising is the large value of $m_s^* \approx
450\  {\rm MeV}$ (the value of the strange quark mass below which the first 
order phase transition is still possible)  as measured in Ref.\cite{Iwas}.
 If it is true, the point
corresponding to our
physical world lies well inside the first order phase
transition region in
which case the first order phase transition should be of
strong variety
which contradicts Ref.\cite{Columb} and also the theoretical
arguments above.

 Again, we do not know what of these two measurements is
more correct.
Possibly, intrinsic lattice artifacts are more dangerous
when an algorithm
with Wilson fermions is used but I 
cannot have an 
opinion on this point. To resolve the controversy, it would
make sense to
perform a {\it quantitative} comparison of the temperature
dependence of
the quark condensate as measured on lattices with the
theoretical curve
in Fig.\ref{Gerberl}. Such a comparison has not been done so
far.

To summarize, at the current level of understanding, the
picture
where the hadron gas goes over
to quark-gluon plasma and other way round without any phase
transition looks more probable. We have instead a sharp
crossover in
a narrow
temperature range which is similar in properties to
second-order phase transition (the ``phase crossover'' if
you
will).

\section{Finite baryon density. Color superconductivity.}
 In real experimental situation, not only the temperature, but also the
baryon number density is high: the colliding nuclei are not only heated,
but also squeezed. What do the theorists know about the properties of
the hadronic system with finite baryon number density ? First of all,
the problem can be solved exactly while the density is still small and
the medium presents a nuclear liquid. In particular, one can derive the
formula describing the suppression the quark condensate in  dense cold
hadronic matter: \cite{Druk} 
  \be
 \label{Druk}
<\bar q q>_n \ =\ <\bar q q>_0 \left[ 1 - \frac {2 n \sigma_{\pi
N}}{F_\pi^2 M_\pi^2} + o(n) \right]
 \ee
Here $\sigma_{\pi N} \approx 45 $MeV is the so called sigma-term
--- the quark contribution to the nucleon mass. The correction turns out
to be surprisingly high. For the nuclear density $n_0 \approx .15 ({\rm
fm})^{-3}$, the drop of the condensate constitutes about $\sim 30$ \% !
One should expect thereby a drastic change in the properties of the
system, probably the phase transition in the region $n \approx 3n_0$.
Basing on the analogy between (\ref{Druk}) and (\ref{qqT}) and the
previous discussion for the finite temperature case, one could guess
that the phase transition is also of the first order.   

This is not true, however, and the phase transition in this case is of
the {\it first} order. Actually, in the imaginary World without
electromagnetic interaction, there would be {\it two} different first
order phase transitions in  density. The existence of the first such
phase transition can be qualitatively explained in a very simple way. It
is well known that, if the protons would not have an electric charge and
would not repel each other, the most favorable energetically state for
the ensemble of a large number of nucleons would be the nuclear matter
state. It has a definite density $n_0$. Suppose we want to study the
system with zero temperature and an overall density $n < n_0$. The
system 
would then present
a set of nuclear matter clusters hovering in the empty space. This is
what is called the {\it mixed} inhomogeneous phase characteristic for
the systems with the first order phase transition (just imagine a
boiling water). The interval of the densities $0 \leq n \leq n_0$ where
the system can exist corresponds to {\it one and the same} value for the
chemical potential $\mu_c$. If we increase the temperature, the nucleons
would evaporate from the clusters and also pion excitations would show
up. While the temperature is small, a system can still exist in mixed
phase, but it would be less inhomogeneous and the interval of allowed net
densities $\Delta n$ for the mixed phase would be smaller than at $T=
0$. At a certain critical temperature $T_c$, the allowed density
interval shrinks to zero and no inhomogenuity in nucleon distribution is
possible if the temperature is still higher. 

But nuclear matter itself is also  an inhomogeneous system: it consists of
nucleons where the density of matter  is higher than in relatively empty
regions of space between them. Qualitatively, one should expect a first
order phase transition associated with squeezing the sytem to the point
that individual nucleons unite in homogeneous clusters carrying large
baryon charge. Again, in some density interval the system would exist in
a mixed phase: dense homogeneous clusters floating in normal nuclear
matter or may be clusters of the rarified nuclear matter immersed in the
dense homogeneous medium, depending on what the net density is. Again,
the density interval where the mixed phase can exist shrinks with
temperature and disappears at some critical point.

Reliable theoretical calculations for the exact position of this second
critical point are not possible now, but reasonable estimates show that
the critical density is $\sim (2.5 - 3.0) n_0$. Model calculations show
that the quark condensate vanishes in the homogeneous phase. They also
show that, at least in the theory with two massless flavors, it is
probably not a critical point, but a {\it tri}critical point, i.e. the
line of first order phase transitions does not end up there but continues
as a line of the second order phase transitions going over to the finite
$T$ second order phase transition at $\mu = 0$. 

 An exciting recent theoretical finding is that, though the standard
quark condensate $<\bar q q>$ vanishes in the high density phase,
another condensate, the colored diquark condensate $<qq>$ which was absent at
sero density, {\it appears}. This phenomenon has been called the color
superconductivity. 

A heuristic physical picture for very high densities
is rather clear. Remember what happens in a usual superconductor. The
electrons on the surface of the Fermi sphere attract each other due to 
phonon exchange and form the Cooper
pairs. This is associated with appearance of the charged order parameter
$<ee>$, Meissner effect, etc. If hadron system is very dense, it makes
no sense to speak of individual hadrons anymore, the system is better
described in terms of quarks 
\footnote{The temperature is still zero, so we have only quarks and no
antiquarks, neither gluons.}.
The colored quarks interact with each other. Consider first their
perturbative interaction due to 1-gluon exchange. In contrast to
electrons which always repel each other, quarks  repel is their color
wave function is symmetric (the diquark is in the sextet representation)
and  attract if it is antisymmetric (the anti-triplet representation).
Thus, there is a way to form Cooper pairs and they are formed.

\begin{figure}
\grpicture{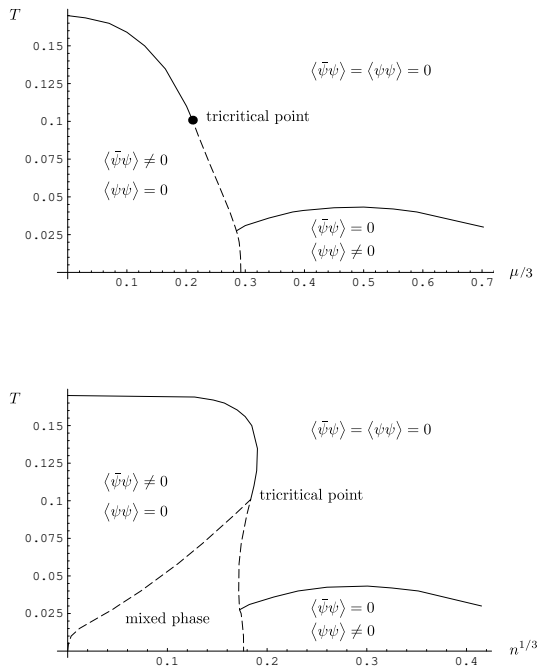}
\caption{Phase diagram of hot and dense QCD.}
\label{Rajplot}
\end{figure}

That was realized already some time ago \cite{Bailin}, but it was
realized also that the perturbative quark interaction is rather weak.
The characteristic values of the superconducting gap were estimated to be
just several MeV which could not lead to any observable effects. An
important recent observation \cite{ShurRaj} was that there is also a
nonperturbative mechanism of forming the Cooper pairs in the attractive
anti-triplet channel due to {\it instantons}. It brings about a rather
large values for the mass gap and for the diquark condensate when the
density is high (above the phase transition), but not asymptotically
high. A possible form of the net phase diagram for the theory with two
massless flavors in the $(T,\mu)$ and $(T, n)$ planes is shown in
Fig.\ref{Rajplot} taken from Ref.\cite{Rajlast}. The solid lines mark
out the second order phase transitions and the dashed lines --- the
first order phase transitions. The line of the trivial first order phase
transition nucleon gas $\to$ nuclear matter is not displayed on these
plots. The values of $T$ and $\mu$ achieved in the real heavy ion
collision experiments are expected to be close to the position of the
upper tricritical point and that may lead to some distinct observable
effects\cite{Rajlast}.

\section{Disoriented chiral condensate.}

I want to discuss in more details here one particular 
 intriguing possibility  that a direct experimental
evidence for the phase transition associated with restoration of chiral
symmetry can be obtained at RHIC.

 After a head-on collision of two energetic heavy
nuclei, a  high temperature hadron ``soup'' is created.
At RHIC
energies, the temperature of the soup would be well above
the
estimate (\ref{est}) for the phase transition temperature.
The
high-temperature state created in heavy nuclei collision
would
exist for a very short time, after which it expands, cools
down and decays eventually into  mesons.

\begin{figure}
\begin{center}
        \epsfxsize=300pt
        \epsfysize=0pt
       \epsfbox[0 320 600 600]{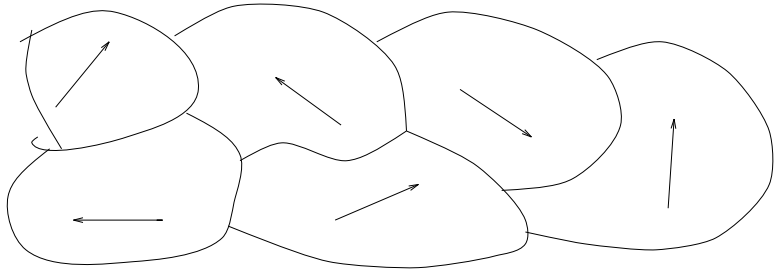}
    \end{center}
\caption{Domains of disoriented chiral condensate in cooling
hadron soup.}
\label{disor}
\end{figure}

Let us look in more details at the cooling stage. At high
temperature, the fermion condensate is zero. Below phase
transition, it is formed and breaks spontaneously chiral
symmetry. This breaking means that the vacuum state is not
invariant under the chiral transformations and a
direction in isotopic space is distinguished. What
particular
direction --- is a matter of chance. This direction is
specified
by the condensate matrix (\ref{Sigff1}).
For simplicity, we have assumed up to now that the
condensate matrix is
diagonal $\Sigma_{ff'} = -\frac \Sigma 2 \delta_{ff'}$.
But any unitary matrix can be substituted for $\delta_{ff'}$
(of course, it can be brought back in the form
$\delta_{ff'}$ by
a chiral transformation ).  In different regions of space,
cooling occurs independently and flavor directions of
condensate are
not correlated. As a result, domains with different
directions
of condensate shown in Fig.\ref{disor} are formed (cf.
cooling down of a
ferromagnet below the Curie point).


In our world, we do not observe any domains, however. The
direction of the condensate in all spatial points is
identical.
This is a consequence of the fact that $u$- and $d$- quarks
have
nonzero masses which break chiral symmetry explicitly, the
vacuum energy involves a term
\be \label{EMSig} E_{vac} \sim
{\rm Tr} \{{\cal M}^\dagger \Sigma \} + {\rm c.c.}
\ee and the
condensate matrix of the true vacuum state is diagonal.

However, the masses of $u$- and $d$- quarks are rather small
and
one can expect that the domains with ``wrong'' direction of
the
condensate are sufficiently developed during the cooling
stage
before they eventually decay into true vacuum
with
emission of pions.

This is a crucial assumption. A theoretic estimate of the
characteristic size of domains they reach before decaying is
very difficult and there is no unique opinion on this
question
in the literature. But if this assumption is true, we can
expect
to observe a very beautiful effect \cite{Bj,Rajag1}.  From the true
vacuum viewpoint, a domain with disoriented $\Sigma_{ff'}$
is a
classical object --- kind of a ``soliton'' (quotation marks
are put
here because it is not stable) presenting a {\it coherent}
superposition of many pions. The mass of this quasi-soliton
is
much larger than the pion mass. The existence of such multi-
pion
coherent states was discussed long ago in pioneer papers
\cite{Ans} but not in relation with thermal phase
transition.

Eventually, these objects decay into pions. Some of the
latter
are neutral and some are charged. As all isotopic
orientations
of the condensate in the domains are equally probable, the
{\it
average} fractions of $\pi^0$, $\pi^+$, and $\pi^-$ are
equal:
$<f_{\pi^0}> = <f_{\pi^\pm}> =
\frac 13$ as is also the case for incoherent production of
pions in, say, $pp$ collisions where no thermalized high-$T$
hadron soup is created.

But the {\it distribution} $P(f)$ over the fraction of, say,
neutral pions is quite different for coherent vs. incoherent
 production. In the incoherent case, $P(f)$ is a very
narrow
Poissonic distribution with the central value $<f_{\pi^0}> =
1/3$. The events with $f_{\pi^0} = 0$ or with $f_{\pi^0} =
1$
are highly unprobable: $P(0) \sim P(1) \sim
\exp\{ - C N \}$ where $N \gg 1$ is the total number
of pions produced.

For coherent production, the picture is quite different.
$\Sigma_{ff'}$ is proportional to a $SU(2)$ matrix.
Factorizing
over $U(1)$, one can define a unit vector in isotopic space
$\in
S^2$. The fraction of $\pi^0$ produced would be just $f =
\cos
^2 \theta$ where $\theta$ is a polar angle on $S^2$.  The
probability to have a particular polar angle $\theta$
normalized
in the interval $0 \leq \theta
\leq \pi/2$ [ the angles $\theta > \pi/2$ do not bring about
anything new as $f(\pi - \theta) = f(\theta)$ ] is
$P(\theta) =
\sin \theta$.  After an elementary transformation, we get a
normalized probability in terms of $f$:
\be \label{Pf}
P(f) df =
\frac {df}{2 \sqrt{f}}
\ee As earlier, $<f> = 1/3$, but the
distribution in $f$ is now wide and the values $f = 0$ and
$f =
1$ are quite probable.

Thus a hope exists that, in experiments with heavy ion
collisions at RHIC, wild fluctuations in the fractions of
neutral and charged pions would be observed. That would be a
direct experimental indication that a quasi-phase-transition
occurs where domains of disoriented chiral condensate of
noticeable size are developed in a cooling stage. One can
recall
in this respect mysterious Centauro events with anomalously
large fraction of neutral or of charged particles observed
in
cosmic ray experiments \cite{Cent}.
Cosmic ray experiment is not a heavy ion collision
experiment
and there are no
good theoretical reasons to expect the formation of hot
hadron
 matter there, but
who knows, may be that still {\it
was} the first experimental observation of the $QCD$ phase
transition ?

\end{document}